\documentclass[lettersize,journal]{IEEEtran}
\usepackage{amsmath,amsfonts}
\usepackage{algorithmic}
\usepackage{algorithm}
\usepackage{array}
\usepackage[caption=false,font=normalsize,labelfont=sf,textfont=sf]{subfig}
\usepackage{textcomp}
\usepackage{stfloats}
\usepackage{url}
\usepackage{verbatim}
\usepackage{graphicx}
\usepackage{cite}
\usepackage{booktabs}
\usepackage{xcolor,colortbl}
\usepackage{multicol}
\usepackage{pbox}
\usepackage{color,soul}
\usepackage{amsmath, bm}
\usepackage{amssymb}
\usepackage{xparse}
\usepackage{array}
\newcolumntype{P}[1]{>{\centering\arraybackslash}p{#1}}
\newcolumntype{M}[1]{>{\centering\arraybackslash}m{#1}}
\usepackage{stackengine}
\newcommand\xrowht[2][0]{\addstackgap[.5\dimexpr#2\relax]{\vphantom{#1}}}
\NewDocumentCommand{\ceil}{s O{} m}{\IfBooleanTF{#1} {\left\lceil#3\right\rceil} {#2\lceil#3#2\rceil} }
\begin{document}
\title{IRS-User Association in IRS-Aided MISO Wireless Networks: Convex Optimization and\\ Machine Learning Approaches}
\author{Hamid Amiriara, Farid Ashtiani, Mahtab Mirmohseni, and Masoumeh Nasiri-Kenari
\thanks{
The authors are with the Department of Electrical Engineering, Sharif University of Technology, Tehran, Iran (email: \{hamid.amiriara, ashtianimt, mirmohseni,
mnasiri\}@sharif.edu)}
\thanks{This work is based upon research funded by Iran National Foundation (INSF) under project No. 4001804.}}
\markboth{Journal of \LaTeX\ Class Files,~Vol.~14, No.~8, November~2022}%
{Shell \MakeLowercase{\textit{et al.}}: A Sample Article Using IEEEtran.cls for IEEE Journals}
\maketitle
\begin{abstract}
This paper concentrates on the problem of associating an intelligent reflecting surface (IRS) to multiple users in a multiple-input single-output (MISO) downlink wireless communication network. The main objective of the paper is to maximize the sum-rate of all users by solving the joint optimization problem of the IRS-user association, IRS reflection, and BS beamforming, formulated as a non-convex mixed-integer optimization problem. The variable separation and relaxation are used to transform the problem into three convex sub-problems, which are alternatively solved through the convex optimization (CO) method. The major drawback of the proposed CO-based algorithm is high computational complexity. Thus, we make use of machine learning (ML) to tackle this problem. To this end, first, we convert the optimization problem into a regression problem. Then, we solve it with feed-forward neural networks (FNNs), trained by CO-based generated data. Simulation results show that the proposed ML-based algorithm has a performance equivalent to the CO-based algorithm, but with less computation complexity due to its offline training procedure.
\end{abstract}
\begin{IEEEkeywords}
Convex optimization (CO), Machine learning (ML), Intelligent reflecting surface (IRS), IRS-user association, Beamforming.
\end{IEEEkeywords}
\section{Introduction}
\IEEEPARstart{W}{ith} the goal of maintaining service for different types of applications, the sixth-generation (6G) wireless communication network (WCN) is aimed to satisfy more stringent requirements than the fifth-generation (5G) in different aspects, such as energy efficiency, ultra-high data rate, global coverage, low latency, and extremely high reliability. Recently both academia and industry are concertedly looking into beyond 5G (B5G) \cite{ref1}. In order to explore the new technologies to make the above requirements feasible intelligent reflecting surface (IRS) is one of the promising technologies to construct the proper infrastructure in 6G networks.

The IRS is a planar surface containing a large number of low-cost passive reflection units (such as cheap printed dipoles), each being capable of independently inducing a controllable phase change on the incident wave \cite{ref33}. By densely placing the IRS in a WCN and intelligently adjusting their reflections, desired propagation characteristics can be achieved by flexibly reconfiguring the signal reflection. This represents a new fundamental approach for improving the capacity and reliability of WCNs by tackling the fading impairment of the channel between the transmitter and the receiver \cite{ref2}.

Although several works have investigated the optimization of the IRS reflection and/or the base station (BS) beamforming coefficients in IRS-aided WCNs, which usually consider a single IRS in their system model \cite{ref3}, recently, some reported works have focused on the general multi-user multi-IRS-aided WCNs to further enhance the system performance \cite{ref8,ref9,ref10,ref11,ref12,ref13}.
While in the meantime, in particular, the IRS can only optimize its passive beamforming for an associated user. This means that each IRS can only effectively serve one associated user by providing tunable channels between the BS and the user.
So, in this practical model, i.e., multi-user multi-IRS aided WCNs, the IRS-user association is a further challenge coupled with two common problems, i.e., determining the IRSs reflection matrix and BS beamforming vector, that should be jointly optimized to improve the system performance and achieve different objectives, such as maximizing the average signal-to-noise ratio \cite{ref8,ref9,ref10}, the received signal power \cite{ref11}, and the sum-rate \cite{ref12,ref13}.
\subsection{Related Works}
In this subsection, we review the existing works on IRS-user association and ML-based WCNs.
\subsubsection{IRS-user association}
In \cite{ref8,ref11}, the authors consider joint optimizing the phase shift parameters of each IRS and the transmit beamforming vector at the BS for a multi-IRS-assisted multiple-input single-output (MISO) system, to maximize the sum-rate and the received signal power, respectively. However, these works have assumed that the IRS-user associations are known and thus did not investigate their optimal design along with this key system issue. Han et al. in \cite{ref12} proposed an iterative complex algorithm named priority-based swapping for solving the mixed-integer non-convex IRS-user association problem to maximize the weighted sum-rate of the IRS-aided heterogeneous networks. The authors in \cite{ref13} investigated the optimization of the IRS reflection vector in a multi-BS cellular system as well as the BS-user association based on an auction algorithm. The problem of assigning the IRS to users was studied in \cite{ref9} which introduces an IRS-user association method to balance the IRS reflection among different BS-user links. The authors in \cite{ref9}, reformulated the IRS-user association problem into an equivalent mixed-integer linear programming (MILP) and proposed an optimal solution based on the branch-and-bound (BB) algorithm. The max-min average signal-to-interference-plus-noise ratio (SINR) problem for multi-user MISO systems with distributed reflecting IRSs was handled and formulated in \cite{ref10} through the design of IRS-user association, employing successive refinement algorithm.

The above iterative algorithms' complexity increases exponentially with the number of IRS reflection units and involves heavy computation demands. Although some optimization techniques such as BB algorithm were proposed to solve the IRS-user association problem with reduced complexity on average, the computational complexity in the worst case is still exponential which makes the solution less attractive for practical implementation.

\subsubsection{ML-based WCNs}
Recently machine learning (ML) approach is investigated for the non-trivial optimization problems that involve extremely high dimensional optimizations such as large-scale multiple-input multiple-output (MIMO) systems with a massive number of array elements \cite{ref4,ref5}, resource allocation of the relay-aided communication with coupled parameters \cite{ref6}, etc.
Likewise, when the number of interactions between the user and the infrastructure increases rapidly due to the large-scale deployment of the IRS in WCNs, the trained ML can be used for significantly reduced complexity and computational time \cite{ref7}.
The application of ML approaches in IRS-aided WCNs has been studied in some works, such as channel estimation \cite{ref14,ref15}, beamforming \cite{ref16}, energy efficiency \cite{ref17}, and security \cite{ref18}.

Khan et al. \cite{ref14} adopted the convolutional neural network (CNN)-based method for estimating the channel matrices of an IRS-aided massive MIMO WCN.
The authors in \cite{ref15} found that in an IRS-aided single-input single-output (SISO) WCN a feed-forward neural network (FNN)-based scheme for estimating the channel state information (CSI) requires lower training overhead than the CNN-based method.
In \cite{ref16}, a two-stage unsupervised learning-based technique for the joint IRS reflection and the BS beamforming optimization in IRS-aided multiuser MISO downlink WCN was developed. The authors demonstrated that the proposed method yields comparable performance to the traditional optimization algorithms while significantly reducing the computational complexity. Liu et al. \cite{ref17} formulated the joint deployment, phase shift design, and power allocation problem in the MISO non-orthogonal multiple access (NOMA) WCN for maximizing the energy efficiency in IRS-assisted platforms. The authors in \cite{ref18} proposed a novel secure beamforming approach based on deep learning for achieving the optimal beamforming policy against eavesdroppers in dynamic environments.

However, to the best of our knowledge, there is no reported research work on assigning the IRS units to the users utilizing the ML approaches to reduce the computational complexity compared to the conventional approaches \cite{ref14,ref15,ref16,ref17,ref18}.

\subsection{Motivations and Contributions of This Work}
As mentioned above, research on jointly optimizing the IRS-user association, IRS reflection, and BS beamforming, in general multi-IRS-aided WCNs is of crucial importance. In addition, current researches on the IRS-user association problem mainly propose iterative algorithms which are not suitable for low-complexity implementation in practice. Using ML approach for this problem is a promising solution, which has not been investigated before.

To fill the above gaps, in this paper, we first derive the rate of each user in a closed-form by taking into account the effects of IRS-user association, BS beamforming, and IRS reflection in multi-user downlink communication in a cellular wireless network, consisting of a multi-antenna BS aided by an IRS with multiple tiles, as shown in Fig.~\ref{figure_1}. Accordingly, we formulate an optimization problem to maximize the sum-rate of all users by jointly optimizing over the IRS-user associations, IRS reflection matrix, and BS beamforming vector.
As the formulated sum-rate maximization problem is a mixed-integer non-convex problem, it is very challenging to be solved in general \cite{ref19}. Thus, we propose a 3-step convex optimization (CO)-based algorithm for alternately solving one of the unknown variables with the other two variables being known. In particular, in step 1 of the algorithm, we obtain an optimal solution for the IRS-user association, with fixed IRS reflection and BS beamforming by relaxed constraints and solving it via the dual method.

To tackle the computational complexity in the above CO-based algorithm, we offer to utilize supervised ML with pre-processing and post-processing techniques to solve the optimization problem. In the pre-processing, an ML is trained by labelled data sets related to CSI. In post-processing, the trained ML is employed to identify the optimal parameters, needless to solve a CO-based algorithm, which can reduce computation latency incurred.

Succinctly, the contributions of this paper are summarized as:
\begin{itemize}{}{}
\item{We formulate the joint IRS-user association, IRS reflection, and BS beamforming optimization problem for a multi-tile IRS-aided downlink WCN.}
\item{We propose a three-step CO-based algorithm to maximize the sum-rate of all users.}
\item{We convert the joint optimization problem to a regression problem to make it solvable by ML approaches.}
\item{Finally, we utilize ML in the joint optimization algorithm to enhance the computational efficiency, which makes the algorithm more attractive for practical implementation.}
\end{itemize}

\subsection{Organization and Notations}
The remainder of the paper is organized as follows. In Section II, we present the IRS structure, system model, and problem formulation. The proposed CO-based algorithm for jointly IRS-user association, IRS reflection, and BS beamforming, to obtain the maximum sum-rate, is discussed in Section III. Section IV presents our proposed ML-based algorithm for solving the optimization problem in an computationally efficient manner. Simulation results and discussions on the network performance are provided in Section V. Finally, we conclude this paper and highlight some future directions in Section VI.

Scalars, vectors, and matrices are denoted by italic letters, bold-face lower-case letters, and bold-face upper-case letters, respectively. ${\mathbb{C}^{n \times m}}$ represents the space of $n \times m$ complex number matrices. For each complex vector $\mathbf{a}$, $\angle \mathbf{a}$ denotes a vector with each element being the phase of the corresponding element in $\mathbf{a}$, $\left\| \mathbf{a} \right\|$ denotes its Euclidean norm, ${\rm{diag}}\left( \mathbf{a} \right)$ denotes a diagonal matrix with each diagonal element being the corresponding element in $\mathbf{a}$,
$\mathbf{A}^{\dagger}$ denotes conjugate transpose of matrix $\mathbf{A}$, ${\log}\left(. \right)$ denotes the logarithm function with base $2$, $\mod[a; b]$ denotes the operator which returns the remainder after division of $a$ (as a dividend) by $b$ (as a divisor),
\ceil[\big]{.} denotes the ceiling function,
${\cal C}{\cal N}\;\left( {a,\sigma^2} \right)$ denotes the distribution of a circularly symmetric complex Gaussian (CSCG) random variable with mean $a$ and variance $\sigma^2$, and $\mathcal{O}(.)$ denotes the Landau's symbol which describe the order of complexity.
\section{System Model and Problem Formulation}
In this section, first, we describe the structure of IRS. It is assumed that the IRS is divided into multiple tiles. The operation of each tile is controlled by a centralized IRS controller that communicates with the BS over a backhaul link. Then, we focus on assigning one IRS tile to a user and derive the rate for the user. Moreover, we present the sum-rate of all users in a cell for the IRS-aided downlink WCN. Exploiting these results, we then develop the sum-rate maximization problem for IRS-assisted WCN, which accounts for the impact of all IRS tiles-users association, the reflection matrix of all tiles, and BS beamforming vector.
\subsection{IRS Structure}
\begin{figure*}[!t]
\centering
\includegraphics[width=4.5in]{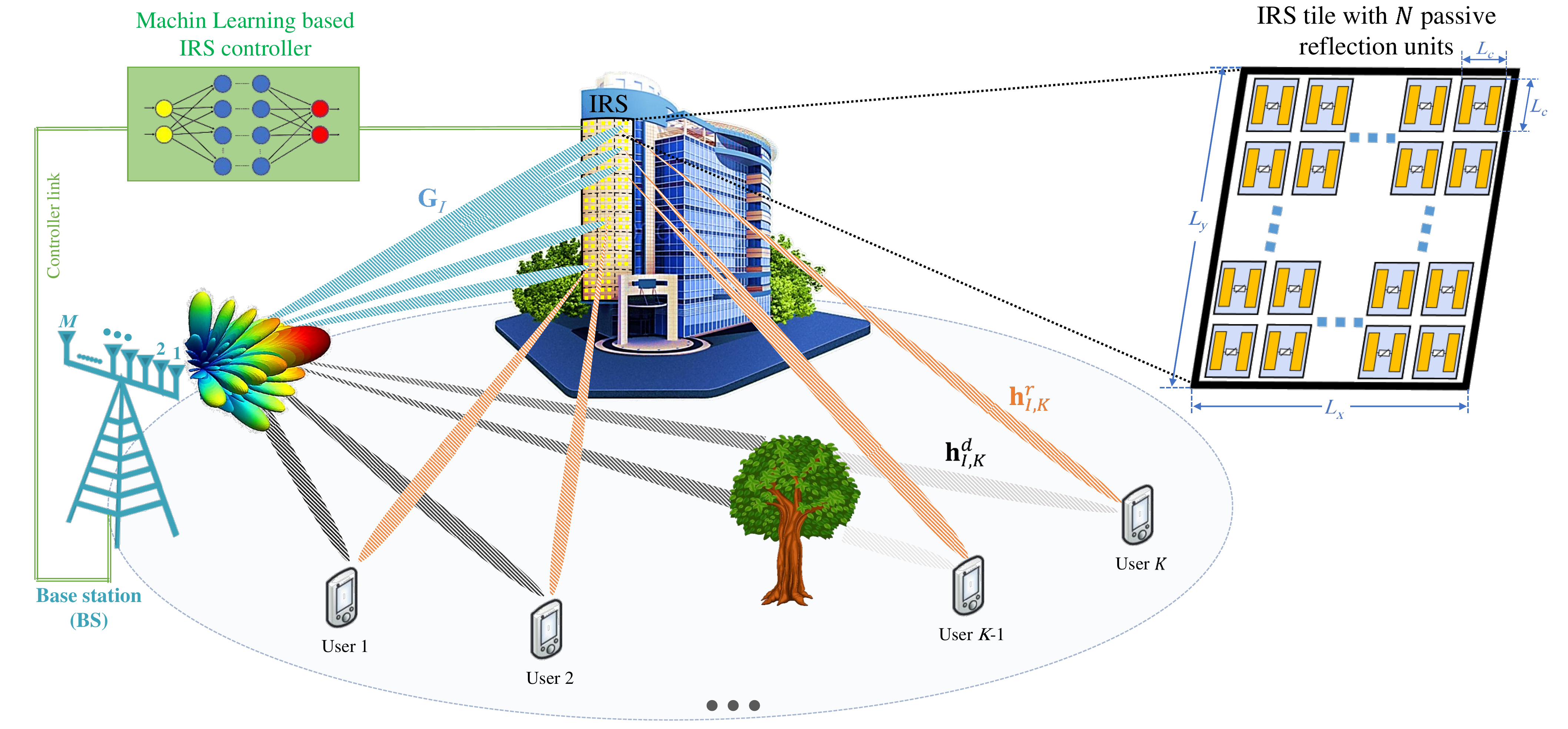}
\caption{MISO downlink wireless communication systems aided by multiple tiles of IRS.}
\label{figure_1}
\end{figure*}

We consider an $x-y$ plane rectangular IRS of size $L_x^{tot}~\times~L_y^{tot}$, see Fig.~\ref{figure_1}. Given a large size for the IRS, it can be partitioned into small tiles of size ${L_x} \times {L_y}$, according to \cite{ref20}. Thus, we assume that there are $I = \;L_x^{tot}L_y^{tot}/\left( {{L_x}{L_y}} \right)$ tiles in total.
Each IRS tile is composed of many sub-wavelength passive reflection units of size ${L_c}\; \times \;{L_c}$ that are able to change the properties of an impinging EM wave when reflecting it. Here, assuming a reflection unit spacing of $dx$ and $dy$ along the $x$ and $y$ axes, respectively, the total number of passive reflection units of each IRS tile is given by $N = {N_x}{N_y}$, where ${N_x} = {L_x}/dx$ and ${N_y} = {L_y}/dy$. When $dx = dy \approx {L_c}\ll\lambda$ and ${L_x},{L_y}\gg\lambda $, the collection of all passive reflection units on one tile acts as a continuous programmable surface \cite{ref18,ref20}.
\subsection{System Model}
As shown in Fig.~\ref{figure_1}, we consider a downlink wireless communication between the BS with $M$ antennas and $K$ single-antenna users. The users are served by the BS and an IRS, composed of $I$ tiles with $N$ passive reflection units. We consider an orthogonal multiple access and so the users are communicated with the BS on orthogonal sub-channels.\footnote{It can be noted that this work can be similarly extended to uplink scenario as well.} For notation simplicity, we denote the sets of users, tiles per IRS, and passive reflection units per tile as ${\cal K} \triangleq \left\{ {1,2, \ldots ,K} \right\}$, ${\cal I} \triangleq\left\{ {1,2, \ldots ,I} \right\}$, and ${\cal N} \triangleq\left\{ {1,2, \ldots ,N} \right\}$, respectively.

Let $\mathbf{h}_k^d \in {\mathbb{C}^{M \times 1}},\;k \in {\cal K}$ be the direct channel coefficient\footnote{As can be seen from Fig.~\ref{figure_1}, the direct channels may be severely in shadow for some users (e.g., blocked by a tree or a building).} from BS to user~$k$, $\mathbf{h}_{i,k}^r \in {\mathbb{C}^{N \times 1}},\;i \in {\cal I},\;k \in {\cal K}$ denote the channel coefficient from the $i$-th IRS tile to user~$k$, and ${\mathbf{G}_i} \in {\mathbb{C}^{N \times M}},i \in {\cal I}$ denote the channel coefficient from BS to the $i$-th IRS tile. In this paper, we consider a rich-scattering propagation environment in the network. Hence, we assume independent Rayleigh fading for all involved channels , and we assume the CSI is perfectly known at BS and IRS, similar to \cite{ref8,ref11}.

To assist in the downlink communications between the BS and users, each user is assumed to be served by some tiles of IRS, such that the user's receiver can constructively combine the directly transmitted signal from the BS with the adjusted reflected signals from the IRS. The binary variables ${\alpha _{i,k}},\;i \in {\cal I},k \in {\cal K}$ are defined, which ${\alpha _{i,k}} = 1$ indicates that the $i$-th tile is associated with the $k$-th user; otherwise, ${\alpha _{i,k}} = 0$.
In order to simplify the IRS-user association and IRS reflection design in practice, the following constraint should be satisfied:
\begin{equation}\label{Equation 1}
\mathop \sum \nolimits_{k \in {\cal K}} {\alpha _{i,k}} = 1,\;\forall i \in {\cal I},
\end{equation}	
which states each tile in ${\cal I}$ can only be associated with at most one user in ${\cal K}$. So, the signal received by the $k$-th user from the BS is written as
\begin{equation}\label{Equation 2}
{y_k} = \left( {\mathop \sum \limits_{i \in {\cal I}} {\alpha _{i,k}}{\mathbf{h}{{_{i,k}^r}^{\dagger}}}{{\mathbf{\Theta }}_i}{\mathbf{G}_i} + \mathbf{h}{{_k^d}^{\dagger}}} \right){{\mathbf{w}}_k}{x_k} + {z_k},
\end{equation}			
where the information signal is represented as ${x_k}$, and ${{\mathbf{w}}_k} \in {\mathbb{C}^{M \times 1}},\;k \in {\cal K}$ denotes the beamforming vector for user~$k$. ${z_k}{{\;}} \sim \;{\cal C}{\cal N}\;\left( {0,\sigma _k^2} \right)$ is the additive white Gaussian noise (AWGN) with zero mean and variance $\sigma _k^2 = {B_k}{N_0}{N_F}$ received at the $k$-th user, where ${B_k}$ represents the bandwidth of each user, ${N_0}$ denotes the noise power spectral density, and ${N_F}$ is the noise figure.

Let the $i$-th tile diagonal IRS reflection matrix be defined by ${{\mathbf{\Theta }}_i} = {\rm{diag}}\left\{ {\left| {{r_{i,1}}} \right|{e^{j{\theta _{i,1}}}},\;\left| {{r_{i,2}}} \right|{e^{j{\theta _{i,2}}}},\; \ldots ,\left| {{r_{i,N}}} \right|{e^{j{\theta _{i,N}}}}} \right\}$, where $\left| {{r_{i,n}}} \right|$ and ${\theta _{i,n}}$ denote the amplitude attenuation and phase shift of the $n$-th reflection unit of tile $i \in {\cal I}$, respectively. In this paper, to achieve the maximum reflected signal power as well as to simplify hardware implementation, it is assumed that the signal is reflected from each tile unit perfectly, i.e., $\left| {{r_{i,n}}} \right| = 1$ \cite{ref21,ref22}.

Thus, the sum-rate of the $k$-th user is given by:
\newcommand\myeq{\mathrel{\overset{\makebox[0pt]{\mbox{\normalfont\tiny\sffamily (a)}}}{\le}}}
\begin{align}\label{Equation 3}
{r_k}= {B_k}\;{\log}\left( {1 + \frac{{{{\left| {\left( {\mathop \sum \nolimits_{i \in {\cal I}} {\alpha _{i,k}}\mathbf{h}{{_{i,k}^r}^{\dagger}}{{\mathbf{\Theta }}_i}{\mathbf{G}_i} + \mathbf{h}{{_k^d}^{\dagger}}} \right){{\mathbf{w}}_k}} \right|}^2}}}{{\sigma _k^2}}} \right).
\end{align}	
Note that in (\ref{Equation 3}), in addition to direct path, $\mathbf{h}{{_k^d}^{\dagger}}{{\mathbf{w}}_k}$, and constructed line of sight (LoS) link (i.e., reflected paths by selected tiles), $\mathop \sum \nolimits_{i \in {\cal I}} {\alpha _{i,k}}\mathbf{h}{{_{i,k}^r}^{\dagger}}{{\mathbf{\Theta }}_i}{\mathbf{G}_i}{{\mathbf{w}}_k} $. In general, other signal paths from the BS to user~$k$ exist (e.g., the interference paths reflected by the non-selected tiles, i.e., $\mathop \sum \nolimits_{i \in {\cal I}} \left(1-{\alpha _{i,k}}\right)\mathbf{h}{{_{i,k}^r}^{\dagger}}{{\mathbf{\Theta }}_i}{\mathbf{G}_i}{{\mathbf{w}}_k} $). However, as will be shown in Fig.~\ref{fig_4}, the gain of these randomly scattered paths can be mitigated to a considerably lower level as compared to constructed LoS link at each user~$k$'s receiver thanks to the optimized BS beamforming vector and IRS reflection matrix designs based on the selected IRS tile in Section III-A and Section III-B thus can be practically ignored \cite{ref9,ref23,ref24}.\footnote{Especially when the pencil-beam condition is valid i.e., the case of practically BS with a massive number of antennas, tiles which have large reflection units, and working in the millimeter-wave bands.}

\subsection{Problem Formulation}
Given the above system model, our goal is to develop a novel algorithm for maximizing the sum-rate of all users in the multiple IRS-aided MISO downlink WCN by jointly optimizing the IRS-user association matrix ${\mathrm{\mathbf{A}}}= \left\{ {{\alpha _{i,k}},\;i \in {\cal I},k \in {\cal K}} \right\}$, IRS reflection matrix ${\mathbf{\Theta }} = \left\{ {{{\mathbf{\Theta }}_i},\;i \in {\cal I}} \right\}$, and BS beamforming vector ${\mathbf{W}} = \left\{ {{\mathbf{w}_k},k \in {\cal K}} \right\}$. Thus, the joint optimization problem is formulated as
\begin{subequations}\label{Equation 4}
\begin{align}
    &\left( {{\rm{P}}} \right):\; \underset{{\bf{{\mathrm{\mathbf{A}}}}},{\mathbf{w}},{\mathbf{\Theta }}}{\max}\mathop \sum \nolimits_{k \in {\cal K}} {r_k}\label{Equation 4A}\\
    & \text{subject to}\;\;\;\;\;\mathop \sum \nolimits_{\;k \in {\cal K}} {\alpha _{i,k}} = 1, \;\;\forall i \in {\cal I}, \label{Equation 4B}\\
    &\;\;\;\;\;\;\;\;\;\;\;\;\;\;\;\;\;\;\;\mathop \sum \nolimits_{\;i \in {\cal I}} {\alpha _{i,k}} = {N_k}, \forall k \in {\cal K}, \label{Equation 4C}\\
    &\;\;\;\;\;\;\;\;\;\;\;\;\;\;\;\;\;\;\;\;{\alpha _{i,k}} \in \left\{ {0,1} \right\},
    \;\;\;\;\;\;\;\forall i \in {\cal I},\forall k \in {\cal K},\label{Equation 4D} \\
    &\;\;\;\;\;\;\;\;\;\;\;\;\;\;\;\;\;\;\;\;{\|\mathbf{w}_k\|}^2 \le {P_t},\;\;\;\;\;\;\;\;\forall k \in {\cal K}, \label{Equation 4E}\\
    &\;\;\;\;\;\;\;\;\;\;\;\;\;\;\;\;\;\;\;\;0 \le {\theta _{i,n}} \le 2\pi ,\;\;\;\;\;\;\forall i \in {\cal I},\forall n \in {\cal N}\label{Equation 4F},
\end{align}
\end{subequations}
where constraint (\ref{Equation 4B}) ensures that each IRS tile is only associated with one user, and ${P_t}$ is the maximum transmit power of the BS for each user. ${N_k}$ denotes the allowed number of associated tiles with one user. If we assume that the free space path-loss of the IRS-aided link is equal to path-loss of the unobstructed direct link, so a minimum of ${N_k}$ can be computed by the following equation \cite{ref25}:

\begin{equation}\label{Equation 5}
{N_{{k}}} = \ceil*[\big]{\frac{{\lambda d_k^td_k^r}}{{{L_x}{L_y}d_k^d}}},
\end{equation}		
where $\lambda$ is the wavelength, and $d_k^d$, $d_k^t$, and $d_k^r$ are the distance of BS-user~$k$, BS-tile~$i$, and tile~$i$-user~$k$ links, respectively.

Note that (P) is a non-convex optimization problem due to the non-concavity of its objective function. Nevertheless, we exploit the special structure of objective function to address this challenge.
Specifically, by the triangle inequality, the rate of the $k$-th user in (\ref{Equation 3}) provides the following inequality:
\begin{align}\label{Equation 3-2}
{r_k} \myeq {B_k}\;{\log}\left( {1 + \frac{\left({\mathop \sum \nolimits_{i \in {\cal I}} {\alpha _{i,k}}\beta _{i,k}^r + \beta _k^d}\right)^2}{{\sigma _k^2}}} \right)\triangleq{r^{\rm{upper}}_k},
\end{align}			
where $\beta _{i,k}^r = {\left| {\mathbf{h}{{_{i,k}^r}^{\dagger}}{{\mathbf{\Theta }}_i}{\mathbf{G}_i}{{\mathbf{w}}_k}} \right|}$ is the gain of the channel between the BS and user~$k$ through being reflected by tile~$i$, $\beta _k^d = {\left| {\mathbf{h}{{_k^d}^{\dagger}}{{\mathbf{w}}_k}} \right|}$ is the direct channel gain between the BS and user~$k$. If $\angle \left( \mathbf{h}{{_{i,k}^r}^{\dagger}}{{\mathbf{\Theta }}_i}{\mathbf{G}_i} {{\mathbf{w}}_k}\right)\; = \angle \left( {\mathbf{h}{{_k^d}^{\dagger}}}{{\mathbf{w}}_k} \right)\triangleq{\bm{\Psi} _k}$, i.e., two additive terms are phase-aligned, inequality (a) in (\ref{Equation 3-2}) holds with equality.
Hence, problem~(P) can be equivalently rewritten as the following problem:
\begin{subequations}\label{Equation 4-2}
\begin{align}
    & \underset{{\bf{{\mathrm{\mathbf{A}}}}},{\mathbf{w}},{\mathbf{\Theta }}}{\max}\mathop \sum \nolimits_{k \in {\cal K}} {r^{\rm{upper}}_k}\label{Equation 4A-2}\\
    & \text{subject to}\;\;\;\;\;\text{(\ref{Equation 4B}), (\ref{Equation 4C}), (\ref{Equation 4D}), (\ref{Equation 4E}), (\ref{Equation 4F}), and}\\
    &\angle \left(\mathbf{h}{{_{i,k}^r}^{\dagger}}{{\mathbf{\Theta }}_i}{\mathbf{G}_i} {{\mathbf{w}}_k}\right)={\bm{\Psi} _k},{\rm{if}}\;{\alpha _{i,k}} = 1, \forall k \in {\cal K},\forall i \in {\cal I} \label{Equation 4G-2}.
\end{align}
\end{subequations}

It can be seen that the considered problem is still a non-convex mixed-integer one, so there is no standard method for obtaining a globally optimal solution for (P). Thus, we propose a 3-step alternating optimization algorithm for solving the problem~(P) approximately based on CO in the following section. In Section IV, the ML-based approach is proposed to achieve a computationally efficient solution for practical and real-time applications.
	
\section{CO-based Approach}
In optimization problem~(P), it is observed that the constraints (\ref{Equation 4B})- (\ref{Equation 4D}) only contain the IRS-user association variable ${\alpha _{i,k}}$, the constraint (\ref{Equation     4E}) only contains the BS beamforming variable ${\mathbf{w}_k}$ and the constraints (\ref{Equation 4F}) only contains the IRS reflection variable ${\theta _n}$. This motivates us to solve the problem~(P) with the alternating optimization (AO) approach. Specifically, we solve it by optimizing ${\alpha _{i,k}}$, ${\mathbf{w}_k}$, and ${\theta _n}$, alternately with the following three related sub-problems:
\newline
 \textit{Sub-problem 1}: Optimization of IRS-user association with given BS beamforming vector and IRS reflection matrix.
 \newline
 \textit{Sub-problem 2}: Optimization of BS beamforming vector for given IRS-user association and IRS reflection matrix.
 \newline
 \textit{Sub-problem 3}: Optimization of IRS reflection matrix with given IRS-user association and BS beamforming vector.
 \newline
Then, we present the overall algorithm and in the last subsection, its convergence and complexity are investigated.
\subsection{Sub-Problem 1: IRS-user Association Optimization}
By fixed beamforming vector ($\mathbf{w}_k$) and IRS reflection matrix (${\mathbf{\Theta }}_i$) we formulate sub-problem 1 in the following form,
\begin{subequations}\label{Equation 6}
\begin{align}
    &\left( {{\rm{P}}1} \right):\; \underset{{\mathrm{{\mathbf{A}}}}}{\max}
     \mathop \sum \nolimits_{k \in {\cal K}} {r^{\rm{upper}}_k} \label{Equation 6A}\\
    & \text{subject to}\;\;\;\;\;
     \text{(\ref{Equation 4B}), (\ref{Equation 4C}), and (\ref{Equation 4D})}.
\end{align}
\end{subequations}

According to integer constraints (\ref{Equation 4D}), (P1) is a linear integer problem that is hard to be solved. We can convert problem~(P1) to a linear convex problem by relaxing (\ref{Equation 4D}) with considering ${\alpha _{i,k}} \in \left[ {0,1} \right]$ instead. Now we can derive the optimal solution of the relaxed problem according to the dual method \cite{ref26}, so its corresponding Lagrange function according to constraint~(\ref{Equation 4B}) can be derived as
\begin{align}\label{Equation 7}
{\cal L}\left( {\mathrm{\mathbf{A}},{\bm{\lambda }}} \right)
= \mathop \sum \nolimits_{k \in {\cal K}} {r^{\rm{upper}}_k}+ \mathop \sum \nolimits_{k \in {\cal K}\;} \mathop \sum \nolimits_{i \in {\cal I}} \left( {{\lambda _i}{\alpha _{i,k}} - {\lambda _i}} \right),
\end{align}	
where $\bm{\lambda}= \left\{ {{\lambda _i},i \in {\cal I}} \right\}$ is the dual variable. Furthermore, the dual problem~(P1) can be obtained as
\begin{equation}\label{Equation 8}
\underset{{\bm{\lambda}}}{\min}
    {\;\;F\left(\bm{\lambda} \right)}
\end{equation}	
where
\begin{subequations}\label{Equation 9}
\begin{align}
    &F\left( \bm{\lambda}  \right) = \underset{\mathrm{\mathbf{A}}}{\max}
    & & {\cal L}\left( {\mathrm{\mathbf{A}},{\bm{\lambda }}} \right) \\
    & \text{subject to}
    && \mathop \sum \nolimits_{\;i \in {\cal I}} {\alpha _{i,k}} = {N_k}, \forall k \in {\cal K}, \\
    &&& {\alpha _{i,k}} \in \left[ {0,1} \right],
    \;\;\;\;\;\;\;\;\;\;\forall i \in {\cal I},\forall k \in {\cal K}.
\end{align}
\end{subequations}

The dual problem~(\ref{Equation 8}) can be solved via Karush-Kuhn-Tucker (KKT) method. From (\ref{Equation 9}), it can be seen that both objective function and constraints in the aforementioned optimization can be decoupled among different users. Thus, the optimal association for user~$k$ can be obtained by solving the following convex problem:
\begin{subequations}\label{Equation 10}
\begin{align}
    &\underset{{\bm{\alpha} _k}}{\max}
     \;\;\;\;\;\;\;{r^{\rm{upper}}_k} + \mathop \sum \nolimits_{i \in {\cal I}} {\lambda _i}{\alpha _{i,k}} \\
    & \text{subject to}\;\;\;\;\mathop \sum \nolimits_{\;i \in {\cal I}} {\alpha _{i,k}} = {N_k}, \\
    & \;\;\;\;\;\;\;\;\;\;\;\;\;\;\;\;\;\;\;{\alpha _{i,k}} \in \left[ {0,1} \right],
    \;\;\;\;\;\;\;\;\;\;\forall i \in {\cal I},
\end{align}
\end{subequations}
where ${\bm{\alpha} _k} = \left[ {{\alpha _{i,k}},i \in {\cal I}} \right] = {\left[ {{\alpha _{1,k}},{\alpha _{2,k}}, \ldots ,{\alpha _{I,k}}} \right]^T}$.

Problem~(\ref{Equation 10}) is convex and its corresponding Lagrangian function can be given as
\begin{align}\label{Equation 11}
\mathop{\cal L'} \left( {{\bm{\alpha}_k},\nu ,\eta ,\xi } \right) &=
{r^{\rm{upper}}_k}+ \mathop \sum \nolimits_{i \in {\cal I}} {\lambda _i}{\alpha _{i,k}}\notag\\
&+ \nu \left( {\mathop \sum \nolimits_{\;i \in {\cal I}} {\alpha _{i,k}} - {N_k}} \right) \notag\\
&+ \mathop \sum \nolimits_{i \in {\cal I}} {\eta _i}{\alpha _{i,k}}\notag\\
 & + \mathop \sum \nolimits_{i \in {\cal I}} {\xi _i}\left( {1 - {\alpha _{i,k}}} \right).
\end{align}		
The KKT conditions of the problem~(\ref{Equation 7}) are
\begin{subequations}\label{Equation 12}
\begin{align}
\frac{{\partial \mathop {\cal L'} \left( {{\bm{\alpha} _k},\nu ,\eta ,\xi } \right)}}{{\partial {\alpha _{i,k}}}}
&= \frac{{\beta _{i,k}^r}}{c_k} + {\lambda _i} + \nu + {\eta _i} - {\xi _i} = 0\label{Equation 12A}\\
\mathop \sum \nolimits_{\;i \in {\cal I}} {\alpha _{i,k}} &= {N_k}\label{Equation 12B}\\
{\eta _i}{\alpha _{i,k}} &= 0\label{Equation 12C}\\
{\xi _i}\left( {1 - {\alpha _{i,k}}} \right) &= 0,\label{Equation 12D}
\end{align}
\end{subequations}
where $c_k$ is defined as
\begin{equation}\label{Equation 13}
c_k = \left( \frac{\ln 2}{2‌B_k} \right)\left( \frac{\sigma _k^2 + \left(\mathop \sum \limits_{i \in {\cal I}} {\alpha _{i,k}}\beta _{i,k}^r + \beta _k^d\right)^2}{\mathop \sum \limits_{i \in {\cal I}} {\alpha _{i,k}}\beta _{i,k}^r + \beta _k^d} \right),k \in {\cal K}.
\end{equation}			

From (\ref{Equation 12A}), (\ref{Equation 12C}), and (\ref{Equation 12D}), it follows that:
\newline
(i) ${\alpha _{i,k}} = 1$ if and only if ${\eta _i} = 0$ and ${\xi _i} > 0$, so $\frac{{\beta _{i,k}^r}}{c_k} + {\lambda _i} > - \nu $.
\newline
(ii) ${\alpha _{i,k}} = 0$ if and only if ${\eta _i} > 0$ and ${\xi _i} = 0$, so $\frac{{\beta _{i,k}^r}}{c_k} + {\lambda _i} < - \nu $.
\newline
Also, equality holds in (\ref{Equation 12B}) if and only if there are only ${N_k}$ tiles such that ${\alpha _{i,k}} = 1$. Therefore, the optimal IRS-user association (i.e., $\alpha _{i,k}^* = 1$) should have the ${N_k}$ highest values of $\frac{{\beta _{i,k}^r}}{c_k} + {\lambda _i}$, i.e., for a typical $k$-th user
\begin{equation}\label{Equation 15}
\alpha _{i,k}^*\left( {c_k,\lambda _i } \right) = \left\{ {\begin{array}{*{20}{l}}
{1,\;\;\;\;\;\rm{if}\; \emph{i} \in \cal A}\\
{0,\;\;\;\;\;\rm{otherwise},}
\end{array}} \right.
\end{equation}		
where $\cal A$ denotes the set of indices of $N_k$ tiles with the highest value of $\beta _{i,k}^r/c_k + {\lambda _i},i\in{\cal I}$.
According to (\ref{Equation 15}), we obtain the optimal $\alpha _{i,k}^*$ as a function of $c_k$ and ${\lambda _i}$.

Substitution of (\ref{Equation 15}) in (\ref{Equation 13}) yields that the right-hand side of (\ref{Equation 13}) is also a function of $c_k$, i.e., $c_k=F(c_k)$, thus we adopt a numerically root-finding algorithm such as the well-known fixed-point iteration method to obtain the unique value of $c_k$.

It remains to calculate the value of $\bm{\lambda} $ using the unconstrained optimization problem~(\ref{Equation 8}), for which a gradient decent method can be employed to decrease the objective function and converge to a stationary point \cite{ref27}. The proposed optimal IRS-user association is summarized in Algorithm~\ref{Algorithm 1}.

\subsection{Sub-Problem 2: BS Beamforming Optimization}
BS beamforming vector optimization problem with given IRS-user association (${\alpha _{i,k}}$) and IRS reflection matrix (${\mathbf{\Theta }}_i$) can be formulated as
\begin{subequations}\label{Equation 16}
\begin{align}
    &\left( {{\rm{P}}2} \right):\; \underset{{\mathbf{w}}}{\max}
     \mathop \sum \nolimits_{k \in {\cal K}} {r^{\rm{upper}}_k}\\
    & \text{subject to}\;\;\;\;
    \text{(\ref{Equation 4E})}.
\end{align}
\end{subequations}

In this paper, we assume that BS perfectly knows all communication channels. BS is interested in maximizing the signal-to-noise ratio (SNR) of each user using a beamforming vector ${{\mathbf{w}}_k}$. That is, the optimal BS beamforming solution in (\ref{Equation 16}) is obtained by the maximum-ratio transmission (MRT) \cite{ref28}
\begin{equation}\label{Equation 17}
\mathbf{w}_k^* = \sqrt {{P_t}}\frac{{\mathop \sum \nolimits_{i \in {\cal I}} {\alpha _{i,k}}\mathbf{G}_i^{\dagger}{\mathbf{\Theta }}_{{i}}^{\dagger}\mathbf{h}_{i,k}^r + \mathbf{h}_k^d}}{\|{{\mathop \sum \nolimits_{i \in {\cal I}} {\alpha _{i,k}}\mathbf{G}_i^{\dagger}{\mathbf{\Theta }}_{{i}}^{\dagger}\mathbf{h}_{i,k}^r + \mathbf{h}_k^d}}\|}
\end{equation}	

From (\ref{Equation 17}), it can be seen that if the channel between the BS and the user, $\mathbf{h}_k^d$, is weak (or blocked) and tile~$i$ is selected for assistance, the BS prefers to form the beam towards tile~$i$ of the IRS. Otherwise, BS forms the beam directly to the user.
\subsection{Sub-Problem 3: IRS Reflection Optimization}
Since we fix the IRS-user association and BS beamforming in Sub-problem 3, (P) is reduced to an optimization problem with IRS reflection only
\begin{subequations}\label{Equation 18}
\begin{align}
    &\left( {{\rm{P}}3} \right):\;\; \underset{\mathbf{\Theta }}{\max}
    \mathop \sum \nolimits_{k \in {\cal K}} {r^{\rm{upper}}_k}\label{Equation 18-a}\\
    & \text{subject to}\;\;\;\;\;
    \text{(\ref{Equation 4F}) and (\ref{Equation     4G-2})}.
\end{align}
\end{subequations}

Let $\mathbf{h}{{_{i,k}^r}^{\dagger}}{{\mathbf{\Theta }}_i}{\mathbf{G}_i}{{\mathbf{w}}_k}=\mathbf{\theta}_i{\mathbf{H} _{i,k}^r}$ where $\mathbf{\theta}_i=\left[{{e^{j{\theta _{i,1}}}},\;{e^{j{\theta _{i,2}}}},\; \ldots ,{e^{j{\theta _{i,N}}}}}\right]$ and ${\mathbf{H} _{i,k}^r}=\rm{diag}\left(\mathbf{h}{{_{i,k}^r}^{\dagger}}\right){\mathbf{G}_i}{{\mathbf{w}}_k}$ denote the equivalent reflected channel matrix. It is obvious that (\ref{Equation 18-a}) is not function of $\mathbf{\Theta }$, so we can easily obtain the optimal solution of (P3) by (\ref{Equation     4G-2}) as $\mathbf{\theta}_i^*={e^{j\left({\bm{\Psi} _k}-\angle\left(\mathbf{H} _{i,k}^r\right)\right)}}$. Thus, in an optimal manner the $n$-th phase shift of tile~$i$ should be set as \cite{ref3,ref29}:
\begin{equation}\label{Equation 19}
\theta _{i,n}^*= \mod\left[ {{\bm{\Psi} _k} - {\Phi _{i,k,n}};2\pi } \right], \rm{if}\;{\alpha _{i,k}} = 1,\forall k \in {\cal K},\forall i \in {\cal I},
\end{equation}	
where ${\bm{\Psi} _k{{\mathbf{w}}_k}} = \angle \left({\mathbf{h}{_k^{d}}}^{\dagger}\right)$ is the phase of the signal received directly from the BS by the $k$-th user. Moreover, ${\Phi _{i,k,n}} = \angle \left({{h}_{i,k}^r{\left( n \right)}^{\dagger}}\mathbf{h}_{i,n}^t{{\mathbf{w}}_k}\right)$ is the phase of the signal reflected by the $n$-th reflection unit of tile~$i$ to the $k$-th user, where ${h}_{i,k}^r\left( n \right)$ is the $n$-th element of vector $\mathbf{h}_{i,k}^r$, and vector $\mathbf{h}_{i,n}^t \in {\mathbb{C}^{1 \times M}}$ is the $n$-th row of matrix ${\mathbf{G}_i} \in {\mathbb{C}^{N \times M}}$.
\begin{algorithm}[t!]
\caption{Proposed algorithm for sub-problem 1 (P1).}
\begin{algorithmic}
\STATE
\STATE 1:  \textbf{Initialize} $\mathbf{\bm{\lambda}}$
\STATE 2:  $\textbf{repeat}$
\STATE 3:\hspace{0.5cm}Calculate $\bm{c}= \left\{ {{c_k},k \in {\cal K}} \right\}$ by solving (\ref{Equation 13}) using the \STATE \hspace{0.8cm} fixed-point iteration method.
\STATE 4:\hspace{0.5cm}Update IRS-user association based on (\ref{Equation 15}).
\STATE 5:\hspace{0.5cm}Update $\bm{\lambda}$ by the gradient decent method.
\STATE 6:  $\textbf{until}$ converges
\STATE 7:  Output: $\mathrm{\mathbf{A}}^*$.
\end{algorithmic}\label{Algorithm 1}
\end{algorithm}
\subsection{Overall Algorithm}
One can observe that problem~(P) is a non-convex mixed-integer optimization problem due to the integer IRS assignment constraints and the non-concavity of its objective function with respect to ${\mathbf{W}}$ and ${\mathbf{\Theta }}$. Nevertheless, in previous subsections it has been shown that if we fix the two optimization variables, problem~(P) is reduced to sub-problems solvable by using CO methods, i.e., (\ref{Equation 15}), (\ref{Equation 17}), and (\ref{Equation 19}), which thus motivates the alternating approach to solve (P) sub-optimally. The overall alternating algorithm for problem~(P) is summarized in Algorithm~2.
\begin{algorithm}[t!]
\caption{Proposed alternating algorithm for problem~(P).}
\begin{algorithmic}
\STATE
\STATE 1:  \textbf{Initialize} ${\mathbf{W}}$, ${\mathbf{\Theta}}$
\STATE 2:  \textbf{repeat }
\STATE \textit{\textbf{IRS-user association }}\textbf{(P1)}
\STATE 3:\hspace{0.5cm}Perform the user association $\mathrm{\mathbf{A}}$ according to Algo-
\STATE\hspace{0.8cm}rithm~\ref{Algorithm 1}.
\STATE \textit{\textbf{Beamforming }}\textbf{(P2)}
\STATE 4:\hspace{0.5cm}Calculate the beamforming vector of BS ${\mathbf{W}}$ based on
\STATE\hspace{0.8cm}(\ref{Equation 17}).
\STATE \textit{\textbf{IRS Reflection optimization }}\textbf{(P3)}
\STATE 5:\hspace{0.5cm}Calculate the phase shift of IRS ${\mathbf{\Theta }}$ based on (\ref{Equation 19}).
\STATE 6:  $\textbf{until}$ converges
\STATE 7:  Output: $\mathrm{\mathbf{A}}^*$, ${\mathbf{W}}^*$, ${\mathbf{\Theta }}^*$.
\end{algorithmic}\label{Algorithm 2}
\end{algorithm}

\subsection{Convergence and Computational Complexity Analysis of CO-based Approach}
In this part, we analyze the convergence and complexity of the overall Co-based algorithm.
\subsubsection{Convergence analysis}Since the optimal solution is obtained for each sub-problem~(P1) to (P3), the sum-rate is non-decreasing over each step of Algorithm~\ref{Algorithm 2}. Furthermore, due to the feasible set of (P), the sum-rate has an upper bound and cannot increase infinitely. From these facts, we ensure that the proposed algorithm converges. Although no global optimality claim can be made, since problem~(P) is not jointly convex, the proposed algorithm can achieve at least a locally optimal solution based on \cite{ref30}.

\subsubsection{Computational complexity analysis}In Algorithm~\ref{Algorithm 2}, the complexity is mainly dominated by the IRS-user association optimization.
Based on the proposed IRS-user association algorithm, i.e., Algorithm~\ref{Algorithm 1}, in each iteration the main complexity is due to steps 3 to 5.
In Step 3, we must calculate $c_k$ by solving (\ref{Equation 13}), using the fixed-point-based search method.
Since in each iteration, we need to compute the $I\times K$ IRS-user association matrix, the complexity of this step can be expressed as $\mathcal{O}\left(KI\log \left(1/\epsilon\right)\right)$, where $\epsilon>0$ is the fixed-point method accuracy.
In Step~4, we update the IRS-user association matrix according to (\ref{Equation 15}), which involves the complexity $\mathcal{O}\left(KI\right)$.
The final part is Step~5, that we calculate the dual variable $\bm{\lambda}$ by gradient descent method, thus the complexity of this procedure is $\mathcal{O}\left(N_{grad}I^2\right)$ where $N_{grad}$ denotes the number of iterations that the gradient method takes to converge.
Meanwhile, define $N_{iter}$ as the maximal number of iterations required for the convergence.
Thereby, the total complexity of Algorithm~\ref{Algorithm 1} is given by $\mathcal{O}\left(N_{iter}\left(KI(1+\log \left(1/\epsilon\right))+N_{grad}I^2\right)\right)$.

\section{ML-based Approach}
In this section, to reduce the computational complexity and enhance the practicality of the CO-based IRS-user association algorithm (Algorithm~\ref{Algorithm 1}), we exploit the supervised ML approach. In this regard, first, the optimization problem is converted to a regression problem, which is then solved by the ML method.
\subsection{Dataset Generation}
As shown in Fig.~\ref{figure_1}, we consider $K$ users and $I$ tiles of IRS. In each independent run, the BS-user (direct path) and BS-IRS-user (reflection path) channel gains, i.e., $\beta _k^d$, $\beta _{i,k}^r$, respectively, are given as inputs. The IRS-user association matrix, i.e., $\mathrm{\mathbf{A}}$, is accordingly calculated by using Algorithm~\ref{Algorithm 1} as a training target. The dataset with the total number of data $S = 10000$ is generated. Then it is split into 3 datasets as follows: $70\%$ and $10\%$ of the data are used as training and testing sets, respectively. $20\%$ of data are held out for validation set which is used to provide a reference for checking the generalization ability of the trained ML.
\subsection{Preprocessing}
The main goal of this part is that the ML methods learn the relationship between the input parameters, i.e., $\mathbf{Input} = \left\{ {\beta _k^d,{{\;}}\beta _{i,k}^r} \right\},i \in {\cal I},k \in {\cal K}$ and the optimal IRS-user association matrix $\mathbf{\mathbf{A}} = \left\{ {{{{\alpha }}_{i,k}}} \right\},i \in~{\cal I},k \in {\cal K}$. For this purpose, without loss of generality, the association matrix ${\mathrm{\mathbf{A}}}$ is
considered as vector ${\boldsymbol{\alpha}}= \left[ {{{\bar \alpha }_1},\; \cdots ,{{\bar \alpha }_K}} \right],\;\forall k \in {\cal K}$ with the entries ${\bar \alpha _k} \in \left[ {0,{2^K} - 1} \right]$, which is the transformation of each column in ${\mathrm{\mathbf{A}}}$, i.e., ${\left[ {{\alpha _{i,1}},{\alpha _{i,2}},\; \cdots ,{\alpha _{i,K}}} \right]^T},\;i \in {\cal I}$, to an integer. So, the optimization problem can be converted to a regression problem, which can be solved by the ML framework. \subsection{ML Model}
\begin{figure}[!t]
\centering
\includegraphics[width=3.5in]{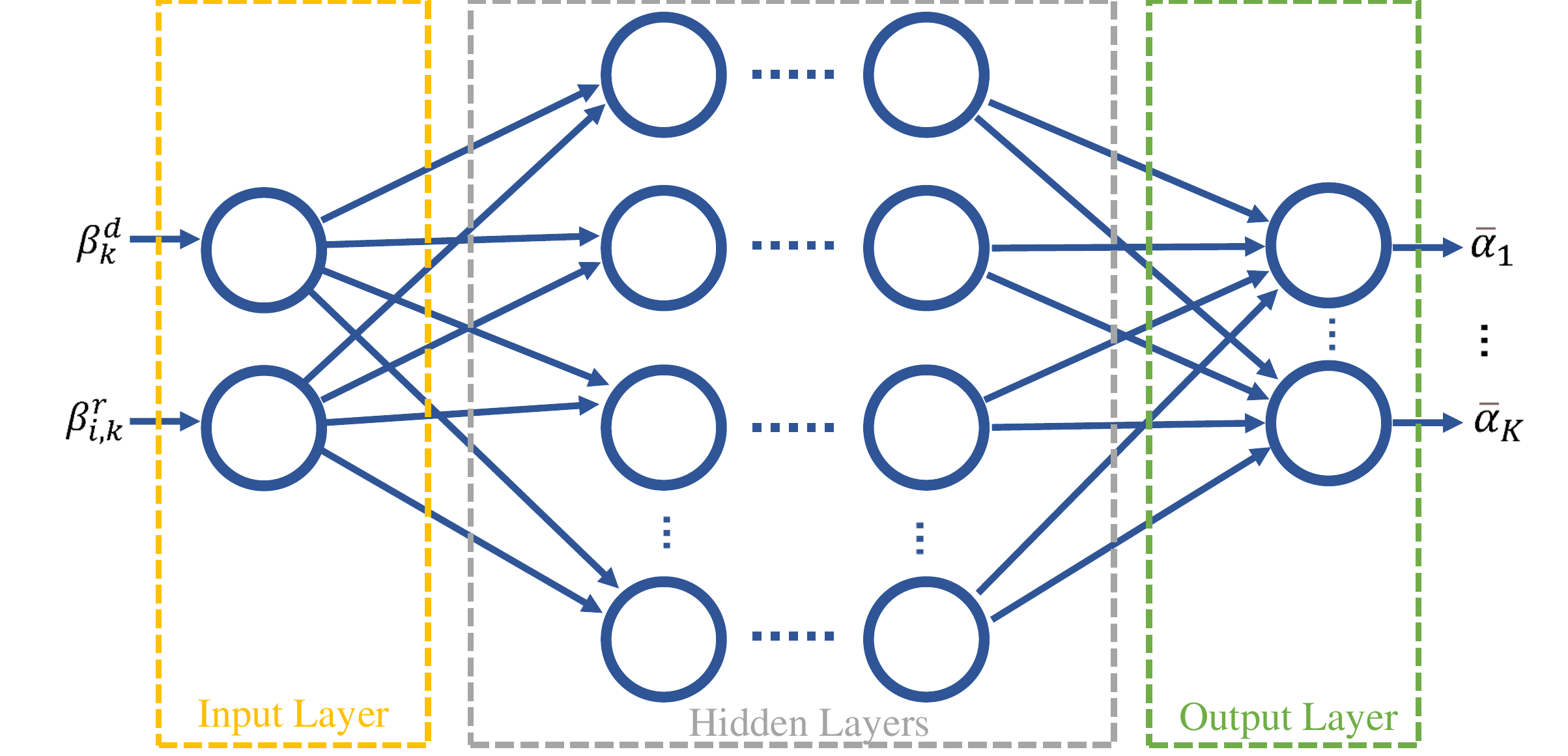}
\caption{The structure of FNNs as ML models used in this paper.}
\label{Figure_2}
\end{figure}

Among the variety of ML approaches found in the literature, suitable for our regression problem, feed-forward neural networks (FNNs) have been adopted for this work. A detailed consideration of neural network structures is not addressed here and is beyond the scope of this work. The configuration of FNNs is also determined by trial-and-error and taking the input size into consideration. The input goes through three hidden layers with ([10, 10, 10], [20, 20, 20]) neurons, and the predicted output (i.e., $\mathrm{\hat{\mathbf {A}}}$) is generated. The loss function is generated by computing the mean squared error of the output value $\mathrm{\hat{\mathbf {A}}}$ and the target value ${{ {\mathrm{\mathbf{A}}}}}$. Then, for each epoch, the optimizer will iteratively optimize the weight values in each layer by Levenberg-Marquardt method \cite{ref31}, based on the loss value. The block diagram of FNN used in this paper is depicted in Fig.~\ref{Figure_2}.

\subsection{Postprocessing}
The outputs of the ML model are in real forms; however, the real output should be changed to integers to determine the exact indices of IRS tiles. Thus, the rounding approach is employed finally to process the outputs of ML. Then, outputs in integer values between $\left[ {0,{2^K}-1} \right]$ are obtained.
\subsection{Benchmark}
In order to provide observations of models' predictive results (i.e., $\mathrm{\hat{\mathbf {A}}}$), the proportion of correctly predicted association is defined as the prediction accuracy. More specifically, assuming that the number of samples is $S$ , the accuracy ($ACC$) is formulated as follows:
\begin{equation}\label{Equation 20}
ACC\left( \%\right) = \frac{{100\% }}{S}\sum \mathbb{I}\left\{ {\mathrm{{\mathbf {A}}} = \mathrm{\hat{\mathbf {A}}}} \right\},
\end{equation}	
where $\mathbb{I}\left\{ . \right\}$ is the Iverson bracket \cite{ref32}.
\subsection{Computational Complexity Analysis of ML-based Approach}
In this section, we propose an ML-based approach for the IRS-user association problem, which is based on the supervised learning algorithm.
During deployment and testing in the ML-based approach each input to the network is only passed through the network once, with a decision at the output layer, so it avoids the iterative operation of fixed-point and gradient descent methods in the CO-based approach, which significantly reduces the computational complexity. The proposed ML-based scheme contains a FNN with $l=3$ number of hidden layers and $n_{l}=10$ or $20$ neurons in each layer. For each layer of the neural network, the computational complexity can be expressed as $\mathcal{O}\left(I_l O_l\right)$, where $I_l$ and $O_l$ are the input and output dimensions of the $l$-th layer, respectively. Hence, the total computational complexity in our ML-based approach is $\mathcal{O}\left((3K+I) n_l+n_l^2\right)$.\footnote{It is worth pointing out that during training, a network is trained over many iterations, so the complexity of the training is the network total computational complexity times the number of ML iterations. More particularly, the training stage needs a powerful computation server that can be performed offline at the ML-based IRS controller while the implementation stage can be completed online.}
So, the ML-based approach diminishes the computational complexity of CO-based IRS-user association from $\mathcal{O}\left(N_{iter}\left(KI(1+\log \left(1/\epsilon\right))+N_{grad}I^2\right)\right)$ to $\mathcal{O}\left((3K+I) n_l+n_l^2\right)$.

\section{Simulation Results}
In this section, some numerical results are provided to evaluate the performance of the proposed algorithms derived in Section III and IV. The channel gain for each effective path (i.e., the elements of ${\mathbf{h}^d}$, ${\mathbf{h}^t}$, and ${\mathbf{h}^r}$ vectors) is modelled as ${h^j} = \sqrt{ {h_{pl}^jh_{ls}^jh_{ss}^j}}, j \in \left\{ {d,t,r} \right\} $, where $h_{pl}^j = {\left( {\frac{\lambda }{{4\pi {d^j}}}} \right)^2}$, $h_{ls}^j$, and $h_{ss}^j \sim \;{\cal C}{\cal N}\;\left( {0,1} \right)$ represent the free-space path-loss, large-scale shadowing/blockage, and small-scale Rayleigh fading, respectively. $d^j, j \in \left\{ {d,t,r} \right\}$ is the link distance where the superscripts $d$, $t$, and $r$ refer to the BS-user, BS-IRS, and IRS-user paths, respectively. It is assumed that the direct links have higher penetration loss due to obstacles in the environment, which motivates the deployment of the IRS.

The stopping threshold for the alternating optimization algorithm is set as $ \epsilon= {10^{ - 4}}$ and the simulation results have been averaged over ${10^3}$ random channel realizations. Other system parameters for BS, IRS, and the users are set as in Table~\ref{Table_1} (if not specified otherwise) \cite{ref25}.

\subsection{Convergence Behavior of the Proposed Alternating Optimization Algorithm}
\begin{figure}[!t]
\centering
\includegraphics[width=3.5in]{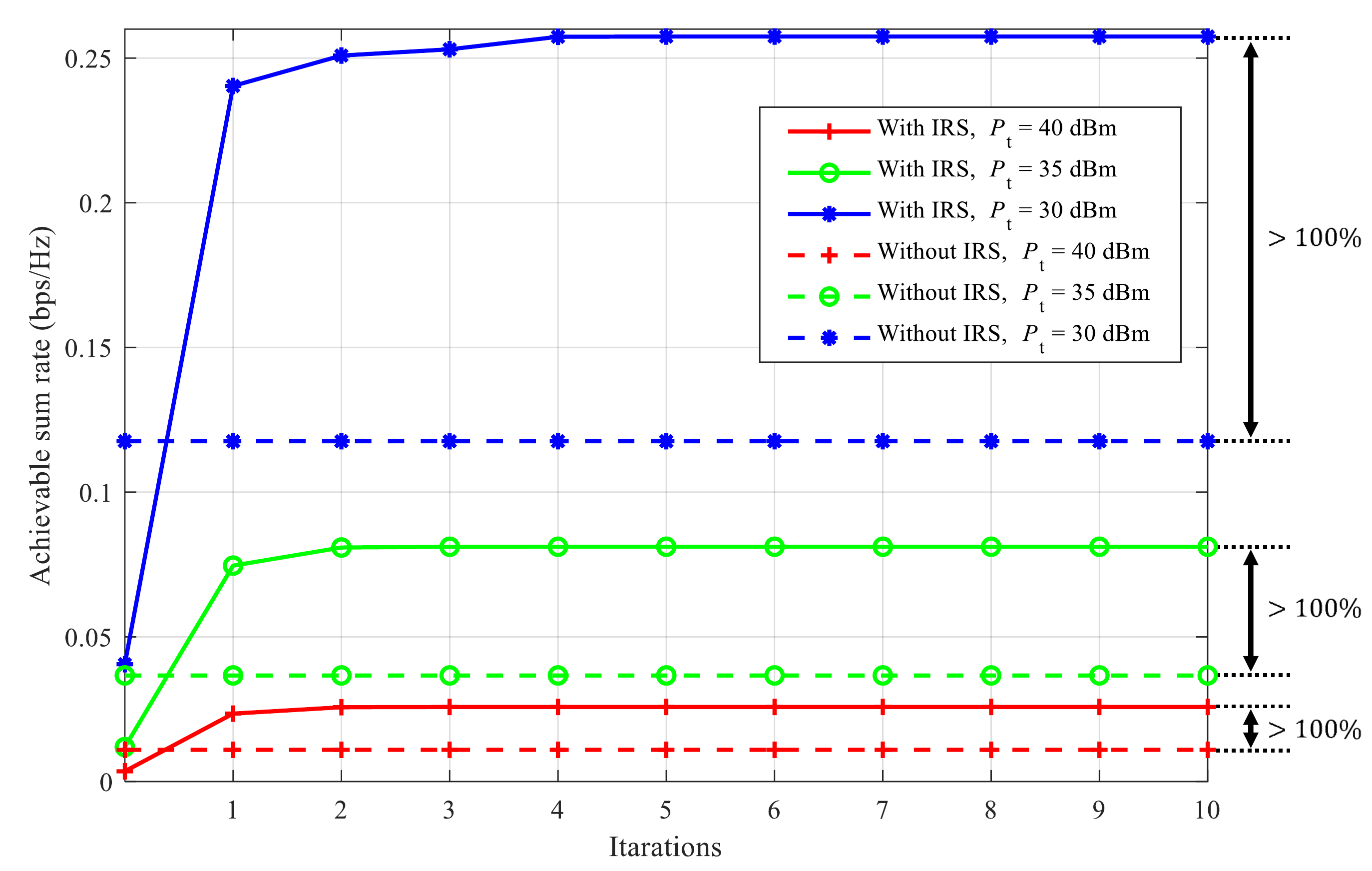}
\caption{Convergence behavior of the proposed alternating optimization algorithm.}
\label{fig_3}
\end{figure}

First, we study the convergence behavior of the proposed 3-step alternating optimization algorithm in Algorithm~\ref{Algorithm 2}. Fig.~\ref{fig_3} shows the achievable sum-rate versus the number of iterations for various transmission powers, i.e., for ${P_t} = 30$, $35$, and $40$~dBm. For each transmission power, the benchmark scheme, in which the IRS is absent but the optimal precoder from sub-problem~(P2) is used by setting ${\mathbf{w}_k} = \sqrt {{P_t}} \frac{{\mathbf{h}_k^d}}{\|{\mathbf{h}_k^d}\|}$, is also shown. As can be observed from Fig.~\ref{fig_3}, Algorithm~\ref{Algorithm 2} converges within 1-4 iterations for different transmission powers usually. We also observe that for the considered channel realization, with the help of Algorithm~\ref{Algorithm 2}, an IRS with one tile of size $10\lambda \times 10\lambda $ can increase the achievable rate by more than $100\% $ compared to the case when an IRS is not employed.
\definecolor{Color1}{rgb}{0.9,1,1}
\definecolor{Color2}{rgb}{0.9,.9,.9}
\renewcommand\arraystretch{1.3}
\begin{table}[!t]
\caption{System parameters for BS, IRS, and users}
\centering
\begin{tabular}{p{39pt}p{127pt}p{50pt}}
\hline
 \rowcolor{Color2}\bf{Symbol} &\bf{Description} &\bf{Value}\\[0.1cm]
\hline\hline
$I$     &Number of tiles        & $4$\\

\rowcolor{Color1}$M$     &Number of BS antennas & $8$\\

$K$     &Number of single-antenna users        & $4$\\

\rowcolor{Color1}${L_x}$, ${L_y}$ &Length of each tile along $x$- and $y$-axes        & $10\;\lambda $\\

${d_x}$, ${d_y}$ &Unit cell spacing along $x$- and $y$-axes& $\lambda /2$\\

\rowcolor{Color1}$N$     &Number of tiles' unit cells & $400$\\

${N_0}$ &Noise power spectral density &$ - 174\;dBm/Hz$\\

\rowcolor{Color1}${N_F}$ &Noise figure & $6\;dB$\\

${B_k},k \in {\cal K}$ &Bandwidth of $k$-th user& $20\;MHz$
\label{Table_1}
\end{tabular}
\end{table}
\subsection{Performance Analysis of the Proposed Algorithm}
\begin{figure}[!t]
\centering
\includegraphics[width=3.5in]{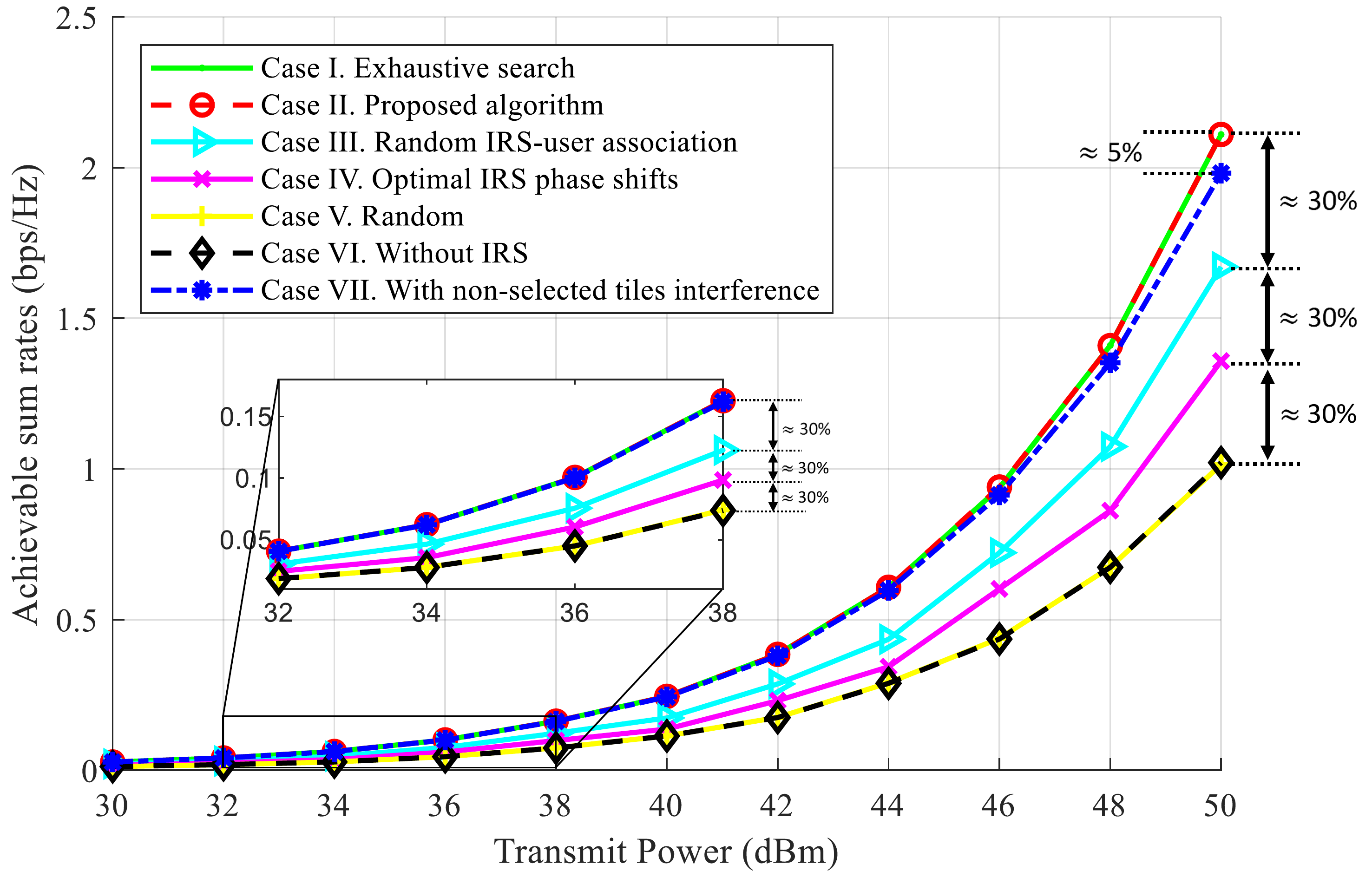}
\caption{Achievable sum-rate versus transmit power (dBm) for different cases.}
\label{fig_4}
\end{figure}

To illustrate the performance of the proposed alternating algorithm, we compare its performance (case II) with the following cases:
\begin{itemize}{}{}
\item{Case I. Exhaustive search: We consider exhaustive search as a benchmark scheme\footnote{The exhaustive search method enumerates all feasible IRS-user associations, which results in a high complexity in the order of $\mathcal{O}(K^I)$. Therefore, it can be considered as a benchmark scheme only for a relatively small system size.}. To this end, at each iteration of Algorithm~\ref{Algorithm 2}, we search all possible IRS-user association (i.e., ${\mathrm{\mathbf{A}}}$) by fixing BS beamforming and IRS reflection matrix, and for each value, we solve (P2) to obtain the optimal beamforming as well as (P3) to optimize the IRS reflection. The above procedure is repeated until convergence satisfied.}
\item{Case III. Random IRS-user association: we set IRS-user association randomly. However, BS Beamforming and IRS reflection are optimized by using the optimal results obtained by solving sub-problems (P2) and (P3), respectively.}
\item{Case IV. Optimal IRS reflection: we set IRS reflection based on (19) but the IRS-user association and beamforming are considered randomly.}
\item{Case V. Random IRS: we set the elements in ${\theta _{i,n}}$ randomly in $\left[ {0,{2\pi}} \right]$ and associate the IRS tiles to users randomly. We set the BS beamforming as ${\mathbf{w}_k} = \sqrt {{P_t}} \frac{{{{\tilde {\mathbf{h}}}_k}}}{\|{{{\tilde {\mathbf{h}}}_k}}\|}$ according to (17).}
\item{Case VI. Without IRS: In this scheme no IRS exists and only beamforming at the BS is applied based on the BS-user direct channel, i.e., ${\mathbf{w}_k} = \sqrt {{P_t}} \frac{{\mathbf{h}_k^d}}{\|{\mathbf{h}_k^d}\|}$.}
\item{Case VII. With non-selected tiles interference: For the purpose of characterizing the effect of the reflect beamforming by associated IRS tiles and random scattering by non-associated IRS tiles on the network performance, the achievable sum-rate while the interference channel gains between the non-selected tiles are not ignored is depicted in this case\footnote{All other parameters are the same as those in case II.}}, i.e., ${r_k}=~{B_k}\;{\log}\left( {1 + \frac{{{{\left| {\left( {\mathop \sum \nolimits_{i \in {\cal I}} {\alpha _{i,k}}\mathbf{h}{{_{i,k}^r}^{\dagger}}{{\mathbf{\Theta }}_i}{\mathbf{G}_i} + \mathbf{h}{{_k^d}^{\dagger}}} \right){{\mathbf{w}}_k}} \right|}^2}}}{\left|{\mathop \sum \nolimits_{i \in {\cal I}} \left(1-{\alpha _{i,k}}\right)\mathbf{h}{{_{i,k}^r}^{\dagger}}{{\mathbf{\Theta }}_i}{\mathbf{G}_i}{{\mathbf{w}}_k}}\right|^2 +{\sigma _k^2}}} \right)$.
\end{itemize}
In Fig.~\ref{fig_4}, we compare the achievable sum-rates for all cases versus the BS's transmission power. First, it is observed that all cases with IRS achieve significant sum-rate enhancement at the users as compared to the case without IRS, except the random case, which demonstrates the effectiveness of IRS-user association and reflecting optimization in IRS.

Fig.~\ref{fig_4} also shows that the proposed alternating algorithm (case II) achieves optimal performance as compared to the exhaustive search (case I), which validates the expression of optimal IRS-user association in Algorithm~\ref{Algorithm 1}. Also, these cases significantly outperform other benchmark schemes, which shows that the joint optimization of IRS-user association, IRS reflection, and BS beamforming are essential to fully reap the gains.

In addition, it is observed that the sum-rate of case IV (i.e., with the optimal IRS reflection, random IRS-user association, and random beamforming) is capable of being increased by almost $30\%$ compared to cases V, VI (which have optimal beamforming). This is because the signal attenuation in the BS-user link is large (i.e., seriously shadowed) and by optimization of the IRS reflection, the users can receive a stronger reflected signal from the IRS-aided link.

Moreover, it can be observed that case III performs approximately $30\%$ better compared with case IV by beamforming the BS optimally. The reason behind such a phenomenon is that in this case, the BS beams are optimally directed towards the IRS and users, which causes the users to fully exploit the potential of beamforming gain.

Furthermore, by comparing cases II and III, it is observed that the performance of the proposed alternating algorithm compared with benchmark schemes is $30\%$ better as the IRS-user association becomes optimized. It is due to the fact that in the former case, the user has received the signal from the optimum links whereas the random links are dominant in the latter case.

Finally, by comparing case VII with case II, it is observed that even with transmitted power $50$ dBm, the achievable sum-rate with considering interference paths reflected by the non-selected tiles (case VII) is about $\%5$ lower than that of the case they are ignored (case II). As transmitted power decreases, this difference decreases, and the former becomes nearly equal to the latter for transmitted power under $42$ dBm. This indicates that the strength of randomly scattered paths in the system is practically much lower than that of the direct paths and constructed LoS link under the optimized BS beamforming vector and IRS reflection matrix designs based on the selected IRS tile and thus can be ignored.
\subsection{Performance Analysis of the ML-based Approach}
In order to verify the validity and computational advantage of the proposed ML-based IRS-user assignment, we present some simulation results in this subsection. For a fair comparison, all the measurements are carried out on some computers with the same configurations. The computation time of both the CO-based algorithm (Algorithm~\ref{Algorithm 1}) and ML-based methods are compared. The overall measurement results are summarized in Table~\ref{Table_2}.
\definecolor{Color1}{rgb}{0.9,1,1}
\definecolor{Color2}{rgb}{0.9,.9,.9}
\begin{table}[!t]
\caption{Computation latency in the case of $10000$ samples using CO-based and ML-based IRS-user association}
\centering
\begin{tabular}{P{48pt}P{20pt}P{20pt}P{23pt}P{22pt}P{23pt}P{22pt}}
\hline
 \rowcolor{Color2}\xrowht{15pt}\bf{Method} &\multicolumn{2}{c}{\bf{Algorithm 1}} &\multicolumn{2}{c}{\pbox{14cm}{\bf{FNN} \\ \bf{(size:[10 10 10])}}} &\multicolumn{2}{c}{\pbox{20cm}{\bf{FNN} \\ \bf{(size:[20 20 20])}}}\\[0.15cm]
\hline\hline

 & Accuracy ($\%$)  & Time (msec)  & Accuracy ($\%$)
 & Time (msec) & Accuracy ($\%$) & Time (msec) \\

\rowcolor{Color1}  $K=2$, $I=2$     &$100$ & $51223$
 &$99.2$ & $42$
 &$99.95$ & $46$\\

 $K=4$, $I=4$     &$100$ & $56452$
 &$98.52$ & $945$
 &$98.92$ & $1026$\\

\rowcolor{Color1}  $K=8$, $I=8$     &$100$ & $59535$
 &$90.02$ & $1705$
 &$94.30$ & $1924$
\label{Table_2}
\end{tabular}
\end{table}

From Table~\ref{Table_2}, it can be seen that the computation time of ML-based approaches is much smaller than the computation time of the CO-based approach, without compromising much prediction accuracy, which implies a considerable computational advantage provided by applying ML for IRS-user association problem. Moreover, we observe that more neurons at the hidden layers lead to a better performance, albeit increasing the computation latency. This shows that the proposed ML-based algorithm can strike a tradeoff between prediction accuracy and computation latency.
\section{Conclusion}
In this paper, the joint optimization of BS beamforming, IRS reflection, and IRS-user association of the multi-tile IRS-aided MISO downlink in multi-user cellular WCN was considered by proposing a three-step CO-based algorithm. The results demonstrated that our proposed algorithm can provide up to $30\% $ and $100\% $ higher sum-rate compared with the random IRS-user association benchmark and absent IRS benchmark, respectively. Additionally, to tackle the computational complexity of the proposed CO-based algorithm in practical implementation, the IRS-user association problem was first converted to a regression problem. Secondly, the machine learning-based method was invoked for determining the control policy of the association of IRS tiles to users. By learning the relation between channel gains and the solution of the CO-based algorithm, the ML-based central IRS controller was shown to be capable of associating the IRS to users in real-time. In particular, the proposed ML-based method could reach a nearly equivalent sum-rate capability with highly less computation time than the CO-based optimization algorithm.

Future studies may extend the theoretical analysis of sum-rate optimization in the presence of the other cells' interferences for multi-cell systems scenarios. Moreover, ML techniques to achieve higher performance in more complicated scenarios deserve more investigation in the future.
\section*{Acknowledgments}
This work is based upon research funded by Iran National Foundation (INSF) under project $No.\;4001804$.

\end{document}